\documentclass[prl,twocolumn,10pt,aps,longbibliography,nofootinbib]{revtex4-1}

 \usepackage{verbatim}
 \usepackage{amsmath}
 \usepackage{amssymb}
 \usepackage{amsthm}
 \usepackage{dsfont}
 \usepackage{latexsym}
 \usepackage{amsfonts}
 \usepackage{epsfig}
 \usepackage{epstopdf}
 \usepackage{color}
 \definecolor{darkblue}{rgb}{0,0,.5}
 \usepackage[linktocpage, colorlinks=true, linkcolor=darkblue, citecolor=darkblue]{hyperref}
 \usepackage[all]{hypcap}

\newcommand{\C}[1]{{\cal{#1}}}
\newcommand{\bb}[1]{\textbf{#1}}

\newcommand{\lr}[1]{{\left\langle {#1}\right\rangle}}

\begin{document}

\title{Response functions as quantifiers of non-Markovianity}

\author{Philipp Strasberg}
\email{philipp.strasberg@uni.lu}
\author{Massimiliano Esposito}
\affiliation{Physics and Materials Science Research unit, University of Luxembourg, L-1511 Luxembourg, Luxembourg}

\date{\today}

\begin{abstract}
 Quantum non-Markovianity is crucially related to the study of dynamical maps, which are usually derived for initially 
 factorized system-bath states. We here demonstrate that linear response theory also provides a way to derive dynamical 
 maps, but for initially correlated (and in general entangled) states. Importantly, these maps are always 
 time-translational invariant and allow for a much simpler quantification of non-Markovianity compared to previous 
 approaches. We apply our theory to the Caldeira-Leggett model, for which our quantifier is valid 
 beyond linear response and can be expressed analytically. We find that a classical Brownian particle coupled to an Ohmic 
 bath can already exhibit non-Markovian behaviour, a phenomenon related to the initial state preparation procedure. 
 Furthermore, for a peaked spectral density we demonstrate that there is no monotonic relation between our quantifier 
 and the system-bath coupling strength, the sharpness of the peak or the resonance frequency in the bath. 
\end{abstract}

\maketitle

\newtheorem*{mydef}{Definition}
\newtheorem{thm}{Theorem}[section]

%%%%%%%%%%%%%%%%%%%%%%%%%%%%%%%%%%%%%%%%%%%%%%%%%%%%%%%%%%%%%%%%%%%%%%%%%%%%%%%%%%%%%%%%%%%%%%%%%%%%%%%%%%%%%%%%%%%%%%%%

{\it Introduction.---} 
A central problem of non-equilibrium statistical mechanics is to obtain a closed dynamical description for some 
``relevant'' degrees of freedom without the need to explicitly model the remaining ``irrelevant'' degrees of freedom. 
Within the theory of open quantum systems, the complete system state $\rho_S(t)$ is usually regarded as relevant while 
the bath is traced out~\cite{BreuerPetruccioneBook2002, DeVegaAlonsoRMP2017}. Using the Nakajima-Zwanzig projection 
operator formalism, this can be done in a formally exact way, but unfortunately, initial system-bath correlations 
prevent the reduced dynamics from being closed due to the appearence of an inhomogeneous term. 

We here show that within linear response theory it is possible (under certain conditions stated below) to obtain 
a reduced dynamical description for a set of system observables even in presence of an initially entangled 
system-bath state. Our findings will allow us to define a rigorous, yet very simple quantifier 
of non-Markovianity, which we can even express analytically for the Caldeira-Leggett model -- a result which is very 
demanding to derive based on previous approaches~\cite{RivasHuelgaPlenioRPP2014, BreuerEtAlRMP2016}. 

{\it Linear response theory.---} 
We consider the standard system-bath setup and assume a global equilibrum state 
$\rho_{SB}(t_0) \sim e^{-\beta(H_S+H_I+H_B)}$ (where $H_{S/I/B}$ denotes the system/interaction/bath part of the
Hamiltonian) prior to the ``initial'' time $t_0$. We then suddenly 
perturb the {\it system part} of the Hamiltonian such that 
\begin{equation}\label{eq perturbed sys Ham}
 H_S(t) = H_S - \sum_{i} a_i\delta(t-t_0)A_i,
\end{equation}
where the $A_i$ are system observables and the $a_i\in\mathbb{R}$ (assumed to be sufficiently small) describe the 
respective strengths of the delta-kick $\delta(t-t_0)$. The purpose of the delta-kick is to generate a local unitary 
transformation $U_0 = \exp(\frac{i}{\hbar}\sum_{i} a_i A_i)\otimes 1_B$, which prepares the system in a nonequilibrium 
state at $t_0$. The initial state has then, to linear order, expectation values 
\begin{equation}
 \lr{A_i(t_0)} = \lr{U_0^\dagger A_i U_0}_\beta = \lr{A_i}_\beta + \sum_j (\chi_+)_{ij} a_j.
\end{equation}
Here, $\lr{\dots}_\beta$ denotes an expectation value with respect to the global equilibrium state. Furthermore, we 
introduced the skew-symmetric matrix $\chi_+$ with entries 
$(\chi_+)_{ij} = \frac{i}{\hbar}\lr{[A_i,A_j]}_\beta$ where $[A,B]$ denotes the commutator. 
We remark that the bath state does not change during this preparation procedure and 
the system-bath correlations (as measured by the mutual information) also remain the same. 

In the following we consider only centered observables such that $\lr{A_i}_\beta = 0$ without loosing generality. 
The expectation value of $A_i$ at a later time $t\ge t_0$ is then connected to the response function 
$\chi_{ij}(t) \equiv \frac{i}{\hbar}\Theta(t)\lr{[A_i(t),A_j]}_\beta$ via the Kubo formula (for an overview 
of linear response theory see Ref.~\cite{PottierBook2010}). In matrix notation we have 
\begin{equation}\label{eq Kubo}
 \lr{\bb A(t)} = \chi(t-t_0) \bb a = G(t-t_0)\lr{\bb A(t_0)},
\end{equation}
where we introduced the mean value propagator $G(t) \equiv \chi(t)\chi_+^{-1}$, which will be the central 
object of interest in what follows. Note that $\lim_{t\searrow 0}\chi(t) = \chi_+$. Eq.~(\ref{eq Kubo}) amounts to our 
fundamental assumption in this paper as it is not guaranteed that the inverse of $\chi_+$ exists 
(this is more thoroughly discussed in the supplementary material). 
If it exists, Eq.~(\ref{eq Kubo}) describes a closed evolution equation for the mean values of the set of observables 
$A_i$ for all times $t\ge t_0$. Two properties of $G(t)$ will be very important in the following. First, $G(t)$ is 
independent of the initial state as it does not depend on any of the $a_i$. Second, the propagator $G(t)$ depends only 
on the elapsed time, which follows from the fact that the response function is expressed in terms of time-translationally 
invariant equilibrium correlation functions (CFs). 

Therefore, if the system behaves Markovian, the mean value propagator must obey 
\begin{equation}\label{eq divisibility}
 G(t) = G(t-s)G(s) ~ \text{ for all } ~ s\in[0,t],
\end{equation}
a condition which is also called divisibility. Equivalently, this implies for the response functions 
\begin{equation}\label{eq time homogeneity}
 \chi(t) = \chi(t-s)\chi_+^{-1}\chi(s). 
\end{equation}

Finally, for later use we note that the response function also determines the temporal behaviour of the equilibrium 
CFs due to the fluctuation dissipation theorem (FDT). Out of the many possible forms of the FDT, we will only need 
\begin{equation}\label{eq correlation FDT}
 \Im[\tilde\chi_{ii}(\omega)] = \frac{1}{2\hbar}(1-e^{-\beta\hbar\omega})\tilde{\C C}_{ii}(\omega),
\end{equation}
where, in general, $\C C_{ij}(t-t_0) \equiv \lr{A_i(t)A_j(t_0)}_\beta$, which also depend only on the time difference. 
Furthermore, we introduced the Fourier transform $\tilde f(\omega) \equiv \int_{-\infty}^\infty dt e^{i\omega\tau} f(\tau)$. 
Note that the FDT also fixes the real part of the response function via the Kramers-Kronig relation. 

{\it Comparison with previous approaches.---} 
Before proceeding, let us contrast our approach with the conventional one. 
Arguably the common considered scenario in the theory of open quantum systems starts with an initial product state 
$\rho_S(t_0)\otimes\rho_B(t_0)$~\cite{BreuerPetruccioneBook2002, DeVegaAlonsoRMP2017, RivasHuelgaPlenioRPP2014, 
BreuerEtAlRMP2016} (for an exception see Ref.~\cite{PollockEtAlPRL2018}). This bears the advantage that the inhomogeneous term in the Nakajima-Zwanzig equation disappears 
and the reduced dynamics of the system is described by a completely positive and trace preserving (CPTP) dynamical map 
\begin{equation}\label{eq CPTP map}
 \Phi(t,t_0)\rho_S(t_0) \equiv \mbox{tr}_B\{U\rho_S(t_0)\otimes\rho_B(t_0)U^\dagger\},
\end{equation}
where $U$ is a unitary evolution operator acting on the joint system-bath state. Although it has been recently studied in 
greater generality whether it is possible to relax the initial product state assumption~\cite{RodriguezRosarioEtAlJPA2008, 
BroducthEtAlPRA2013, BuscemiPRL2014, DominyShabaniLidarQIP2016}, the family of initially entangled states considered above 
will in general {\it not} give rise to a CPTP map. Therefore, there is no direct connection between our approach and 
previous results, although we can draw analogies.  

Indeed, while $\Phi(t,t_0)$ as $G(t)$ is independent of the initial system state, the former does not propagate mean 
values but the complete system state $\rho_S(t_0)$ to arbitrary later times $t\ge t_0$. 
It is interesting to ask whether $G(t)$ can be extended to a dynamical map for the entire system density matrix 
by looking at a complete set of system observables $\{A_i\}$, whose expectation values are isomorphic to $\rho_S(t)$. 
In the supplementary material we demonstrate  that this is not possible because $\chi_+$ in Eq.~(\ref{eq Kubo}) becomes 
non-invertible. 

Finally, to characterize non-Markovianity within the standard approach based on Eq.~(\ref{eq CPTP map}), the concept of 
CP divisibility is important. A CP divisible quantum stochastic process is characterized by a family 
$\{\Phi(t_2,t_1)|t_2\ge t_1\ge t_0\}$ of CPTP maps, which obeys 
\begin{equation}\label{eq CP divisibility}
 \Phi(t_2,t_0) = \Phi(t_2,t_1)\Phi(t_1,t_0) ~ \text{ for all } ~ t_2\ge t_1\ge t_0,
\end{equation}
analogous to the classical Chapman-Kolmogorov equation. Consequently, if a process is CP divisible, then the evolution of 
the density operator is Markovian (although there seems to be less agreement about the converse 
statement~\cite{RivasHuelgaPlenioRPP2014, BreuerEtAlRMP2016}). Based on this concept or a related notion, various 
quantifiers of non-Markovianity have been recently put forward~\cite{WolfEtAlPRL2008, 
BreuerLainePiiloPRL2009, RivasHuelgaPlenioPRL2010, LorenzoPlastinaPaternostroPRA2013, HallEtAlPRA2014, 
ChruscinskiManiscalcoPRL2014} and direct experimental evidence is also accumulating~\cite{LiuEtAlNatPhys2011, 
GroblacherEtAlNatComm2015}. 

Unfortunately, evaluating non-Markovianity for time evolutions generated by Eq.~(\ref{eq CPTP map}) is 
demanding as it requires, e.g., optimization procedures, the inversion of dynamical maps or the integration over 
complicated disconnected domains. In part, this problem is caused by the fact that the CPTP map $\Phi(t,t_0)$ has a 
complicated time dependence: even if the unitary $U$ in Eq.~(\ref{eq CPTP map}) is generated by a time-independent 
Hamiltonian, the dynamical map is not time-translational invariant, i.e., $\Phi(t,t_0) \neq \Phi(t-t_0)$ in general. 
This is in strong contrast to our result in the linear response regime, where $G(t)$ always depends only on the elapsed 
time and which allows us to check the simpler condition~(\ref{eq divisibility}) instead of 
Eq.~(\ref{eq CP divisibility}). 

{\it Distance quantifier.---} To introduce new quantifiers of non-Markovianity within our 
approach, we need to quantify the distance between two functions $f(t)$ and $g(t)$. We use the 
standard $L_2$ scalar product $\lr{f,g} = \int_{-\infty}^\infty dt f(t)g^*(t)$ and the induced norm 
$\|f\| = \sqrt{\lr{f,f}}$, where it is tacitly assumed that the integrals are converging. We then define the distance 
\begin{equation}\label{eq distance f g}
 \C D(f,g) \equiv \sqrt{1-\frac{|\lr{f,g}|^2}{\|f\|^2\|g\|^2}}.
\end{equation}
By Cauchy-Schwarz' inequality $0\le \C D(f,g)\le 1$ and $\C D(\lambda f,\lambda g) = \C D(f,g)$ for any 
$\lambda\in\mathbb{C}$, i.e., the difference has the favourable properties that it is positive, bounded and independent 
of any global scaling. By analogy with the Euclidean scalar product, $\C D(f,g) = |\sin(\phi)|$ can be seen as quantifying 
the ``angle'' $\phi$ between the two vectors $f(t)$ and $g(t)$. Most importantly for our applications, by Parseval's 
theorem we can deduce that $\C D(f,g) = \C D(\tilde f,\tilde g)$, where the right hand side is computed by using the 
$L_2$ scalar product in Fourier space, 
$\langle\tilde f,\tilde g\rangle = \int_{-\infty}^\infty \frac{d\omega}{2\pi} \tilde f(\omega)\tilde g^*(\omega)$. 

{\it New quantifiers of non-Markovianity.---} 
It will be advantageous to work in Fourier space in the following. 
In the supplementary material we show that integrating Eq.~(\ref{eq time homogeneity}) over $s$ from zero to $t$ 
implies in Fourier space 
\begin{equation}\label{eq time homo Fourier}
 -i\frac{d}{d\omega}\tilde\chi(\omega) = \tilde\chi(\omega)\chi_+^{-1}\tilde\chi(\omega). 
\end{equation}
Then, to measure violations of Eq.~(\ref{eq time homo Fourier}) as a consequence of the (assumed) divisibility property, 
we propose the quantifier [denoting $\tilde\chi'(\omega) = \frac{d}{d\omega}\tilde\chi(\omega)$] 
\begin{equation}\label{eq quantifier 1}
 \C N^{(1)}_{ij} \equiv \C D\big[-i\tilde\chi'_{ij},\big(\tilde\chi\chi_+^{-1}\tilde\chi\big)_{ij}\big].
\end{equation}

As a second quantifier of non-Markovianity, we also check the validity of the regression theorem (RT)~\cite{LaxPR1966, 
LaxPR1968}, which allows us to relate the evolution of CFs to the evolution of mean values. Within our setting the 
Markovian assumption enters here by using that 
Eq.~(\ref{eq Kubo}) holds for all initial states and that there exists a dynamical map $\Phi(t,t_0)$, which is 
{\it independent} of $\rho_S(t_0)$. It is worth emphasizing that the validity of the RT 
does not {\it a priori} rely on an initial product state assumption or on the property of CP divisibility. 
It merely signifies that it is possible to find for any initial system state a map $G(t)$ to propagate the mean values 
{\it and} -- in addition to what is required to evaluate $\C N^{(1)}_{ij}$ -- a map $\Phi(t,t_0)$ to propagate 
$\rho_S(t_0)$ (see supplementary material for more details). Thus, if the RT holds 
\begin{equation}\label{eq RT}
 C^\text{RT}(t,t_0) = G(t-t_0)C(t_0,t_0).
\end{equation}
Here, we have added the superscript ``RT'' to emphasize that this is the predicted CF assuming the validity of the RT. 
Note that $C^\text{RT}_{ij}(t,t_0) \equiv \lr{A_i(t)A_j(t_0)}$ denotes in general an out-of-equilibrium CF, but we will 
be only interested in equilibrium CFs which we denote with a calligraphic $\C C$. For them we can deduce in Fourier space 
that (see supplementary material) 
\begin{equation}\label{eq CF prediction RT general}
 \tilde{\C C}^\text{RT}(\omega) = \tilde\chi(\omega)\chi_+^{-1}\C C(0) - \C C(0)^T\chi_+^{-1}\tilde\chi(\omega)^\dagger.
\end{equation}
We add that the behaviour of CFs (often in relation with the validity of the RT) has played an important role historically 
to define a quantum Markov process~\cite{LaxPR1966, LaxPR1968, HaakeBook1973, LindbladCMP1979, TalknerZPhysB1981} and 
was also investigated in the recent debate about non-Markovianity in Refs.~\cite{GuarnieriSmirneVacchiniPRA2014, 
LoGulloEtAlArXiv2014, AliEtAlPRA2015}. However, its use in the linear response regime has not been noted before, although 
it is well-known that all quantum systems violate the RT in that regime~\cite{FordOConnellPRL1996}. 

Then, based on the comparison of the exact equilibrium CFs [obtained from the FDT~(\ref{eq correlation FDT})] and their 
Markovian prediction [obtained from the RT~(\ref{eq CF prediction RT general})], we propose 
\begin{equation}\label{eq quantifier 2}
 \C N_{ij}^{(2)} \equiv \C D\left[\tilde{\C C}_{ij},\tilde{\C C}^\text{RT}_{ij}\right].
\end{equation}
We here assume that the equilibrium covariance matrix $\C C(0)$ is exactly known such that the 
prediction~(\ref{eq CF prediction RT general}) uses the correct initial value.

To conclude, the magnitude of both, $\C N_{ij}^{(1)}$ and $\C N_{ij}^{(2)}$, measures by how much we fail by naively 
assuming that the process is Markovian. They can be computed {\it without} the need to a priori derive any quantum 
master equation -- only the knowledge of the linear response functions {\it or} the equilibrium CFs is required. 

We will now treat an important class of open system models with Gaussian dynamics \emph{exactly}, i.e., without any 
approximation about the temperature of the bath, the system-bath coupling strength or the spectral features of the bath. 
We also remark that for this class our results are valid beyond linear response. Related studies about non-Markovianity 
of Gaussian dynamics based on different approaches and various approximations can be found in 
Refs.~\cite{LorenzoPlastinaPaternostroPRA2013, SouzaEtAlPRA2015, TorreRogaIlluminatiPRL2015, GroblacherEtAlNatComm2015, 
JinYuNJP2018, TorreIlluminatiArXiv2018}. 

{\it Quantum Brownian motion.---} We consider the standard Caldeira-Leggett model with Hamiltonian 
(in suitable mass-weighted coordinates) 
\begin{equation}\label{eq Hamiltonian}
 H = \frac{p^2 + \omega_0^2q^2}{2} + \frac{1}{2}\sum_k\left[p_k^2 + \omega_k^2\left(q_k - \frac{c_k}{\omega_k^2}q\right)^2\right].
\end{equation}
Here, $q$ and $p$ refer to the position and momentum of the system with frequency $\omega_0 > 0$, whereas 
the bath oscillators with frequencies $\omega_k > 0$ are specified with an additional index $k$. Furthermore, $c_k$ 
denotes the coupling strength between the system and the $k$'th oscillator. Of central importance is the spectral density 
(SD) 
\begin{equation}\label{eq SD}
 J(\omega) \equiv \frac{\pi}{2}\sum_k \frac{c_k^2}{\omega_k} \delta(\omega-\omega_k).
\end{equation}
It characterizes the coupling between system and bath and it is assumed to be a continuous function of $\omega$ in 
the limit of a large bath fulfilling $J(0) = 0 = J(\omega\rightarrow\infty)$. A great benefit of the Brownian motion 
model is that almost all quantities of interest are computable in closed form~\cite{GrabertSchrammIngoldPR1988, 
WeissBook2008}, e.g., the matrix of response functions reads (see supplementary material for a derivation) 
\begin{align}\label{eq response fct Brownian}
 \left(\begin{array}{cc}
        \tilde\chi_{qq}(\omega)	&	\tilde\chi_{qp}(\omega)	\\
        \tilde\chi_{pq}(\omega)	&	\tilde\chi_{pp}(\omega)	\\
       \end{array}\right) 
 &= \left(\begin{array}{cc}
            1		&	i\omega	\\
           -i\omega	&	\omega^2	\\
          \end{array}\right)\tilde\chi_{qq}(\omega)
 + \left(\begin{array}{cc}
           0 & 0 \\
           0 & 1 \\
          \end{array}\right),	\nonumber	\\
 \tilde\chi_{qq}(\omega)	&=	\frac{1}{\omega_0^2 - \omega^2 - i\omega\tilde\gamma(\omega)},
\end{align}
where $\tilde\gamma(\omega)$ is the Fourier transform of the memory kernel 
$\gamma(t) = \Theta(t) \frac{2}{\pi}\int_0^\infty d\omega \frac{J(\omega)}{\omega}\cos(\omega t)$. In view of the general 
theory outlined above, our set of system observables will be the position and momentum of the system, 
$\{A_1,A_2\} = \{q,p\}$, and one easily verifies that 
$\chi_+$ is symplectic with $(\chi_+)_{pq} = 1 = -(\chi_+)_{qp}$. The delta-kick now creates the unitary 
\begin{equation}\label{eq unitary preparation}
 U_0 = e^{\frac{i}{\hbar}(a_q q+ a_p p)} = e^{\frac{i}{\hbar}\frac{a_q a_p}{2}} e^{\frac{i}{\hbar}a_q q} e^{\frac{i}{\hbar}a_p p},
\end{equation}
which shifts the position and momentum operators, 
\begin{equation}\label{eq shift q p}
 U_0^\dagger q U_0 = q - a_p, ~~~ U_0^\dagger p U_0 = p + a_q,
\end{equation}
thereby shifting the mean values but leaving the covariances unchanged. 
Furthermore, the equilibrium covariance matrix is diagonal with entries 
$\C C_{qq}(0) = \frac{\hbar}{\pi}\int_0^\infty d\omega \coth(\beta\hbar\omega/2)\Im[\tilde\chi_{qq}(\omega)]$ and 
$\C C_{pp}(0) = \frac{\hbar}{\pi}\int_0^\infty d\omega \omega^2\coth(\beta\hbar\omega/2)\Im[\tilde\chi_{qq}(\omega)]$ 
and the equilibrium CFs are linked via 
\begin{equation}\label{eq relation eq CFs}
 \left(\begin{array}{cc}
        \tilde{\C C}_{qq}(\omega)	&	\tilde{\C C}_{qp}(\omega)	\\
        \tilde{\C C}_{pq}(\omega)	&	\tilde{\C C}_{pp}(\omega)	\\
       \end{array}\right) 
 = \left(\begin{array}{cc}
          1		&	i\omega		\\
          -i\omega	&	\omega^2	\\
         \end{array}\right) \tilde{\C C}_{qq}(\omega). 
\end{equation}
We now have all quantities at hand to compute our quantifiers. For the rest of the paper we will set $t_0 = 0$. 

\begin{figure}%[b]
 \centering\includegraphics[width=0.40\textwidth,clip=true]{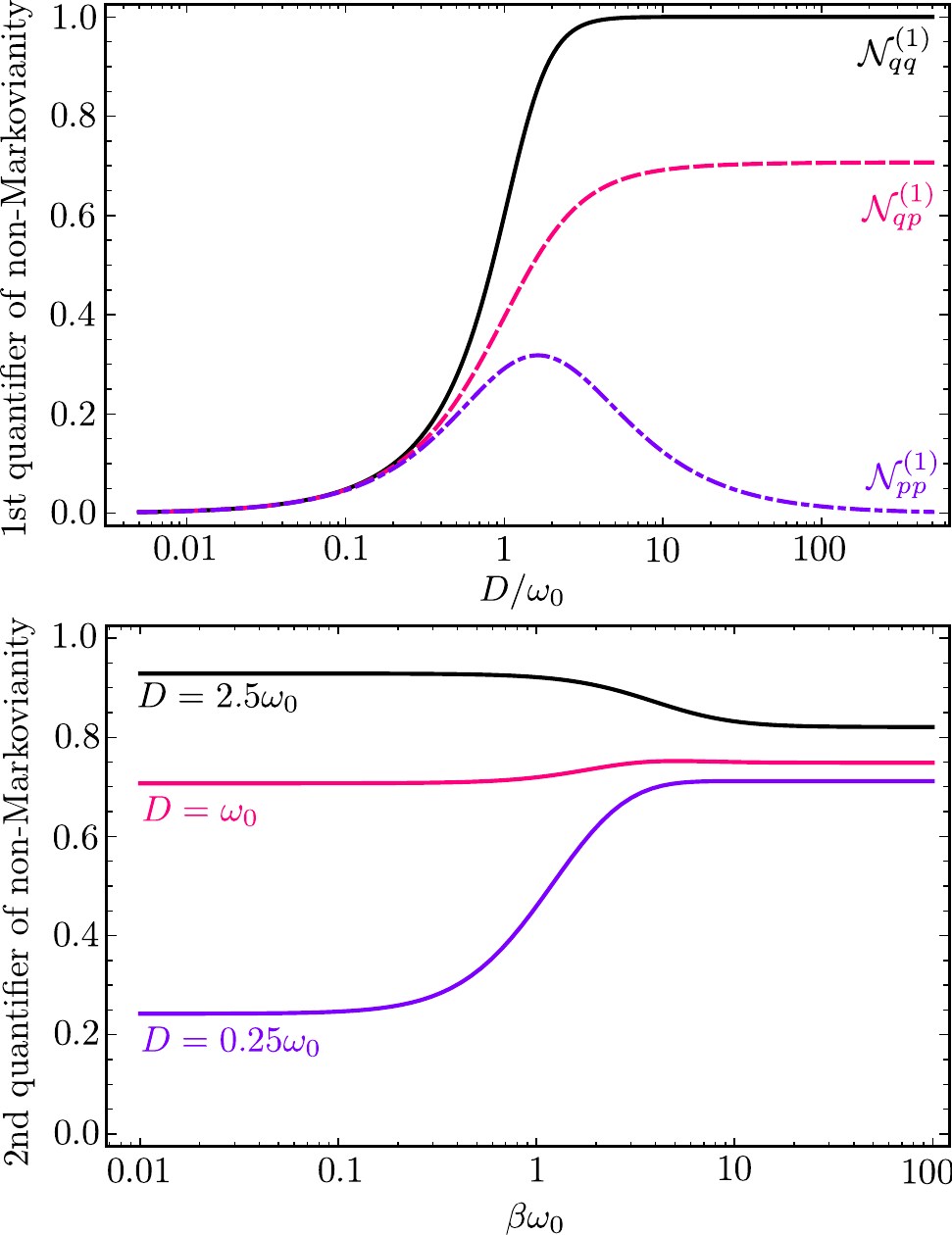}
 \label{fig Ohmic} 
 \caption{\bb{Top:} Plot of our first quantifier of non-Markovianity ($\C N_{qq}^{(1)}$ solid black, $\C N_{qp}^{(1)}$ 
 dashed pink, $\C N_{pp}^{(1)}$ dash-dotted purple) for an Ohmic SD over the dimensionless coupling strength $D/\omega_0$ 
 in logarithmic scale. Note that $\C N_{qp}^{(1)} = \C N_{pq}^{(1)}$. \bb{Bottom:} Plot of the second quantifier 
 $\C N_{qp}^{(2)}$ over the dimensionless inverse temperature $\beta\omega_0$ for various coupling strengths $D$ in 
 logarithmic scale. The other two quantifiers $\C N_{qq}^{(2)}$ and $\C N_{pp}^{(2)}$ (not shown) are similar to the plot 
 in Fig.~\ref{fig plot peaked SD} (right). We set $\hbar\equiv 1$. }
\end{figure}

{\it Classical Ohmic limit.---} We consider the simplest case of a classical particle ($\hbar=0$) coupled to an Ohmic 
bath, which corresponds to a memory kernel of the form $\gamma(t) = D\delta(t)$. This follows from a linear SD 
$J(\omega) = D\omega$ in the limit of an infinitely high cutoff frequency. The resulting Langevin equation for the system 
reads (see supplementary material for a detailed derivation)
\begin{equation}
 \begin{split}\label{eq Langevin anormal}
  \dot q(t)	&=	p(t) - a_p \delta(t),	\\
  \dot p(t)	&=	-\omega_0^2 q(t) + a_q \delta(t) - D\dot q(t) + \xi(t).
 \end{split}
\end{equation}
Here, the noise obeys $\lr{\lr{\xi(t)}} = 0$ and $\lr{\lr{\xi(t)\xi(s)}} = \gamma(t-s)/\beta$ with the crucial requirement 
that $\lr{\lr{\dots}}$ refers to an average over an initial {\it conditional} equilibrium state of the 
bath~\cite{WeissBook2008, BezZPB1980, SanchezCanizaresSolsPhysA1994, HaenggiLectNotes1997} 
\begin{equation}\label{eq cond eq bath}
 \rho_B(0) \sim \exp\left\{-\frac{\beta}{2}\sum_k\left[p_k^2 + \omega_k^2\left(q_k - \frac{c_k}{\omega_k^2}q(0^-)\right)^2\right]\right\}.
\end{equation}
Here, the position $q(0^-)$ of the Brownian particle {\it prior} to the delta-kick is a random variable distributed 
according to a Gaussian $P[q(0^-)] \sim e^{-\beta\omega_0^2q(0^-)^2/2}$ such that, shortly before the unitary kick, the global system-bath 
state is in equilibrium. 

If we would not disturb the state, $a_q = 0$ and $a_p = 0$, and Eq.~(\ref{eq Langevin anormal}) reduces to the standard 
Langevin equation. However, the presence of the unitary kick results is an initial system state described by a shifted 
Gaussian $P[q(0)] \sim e^{-\beta\omega_0^2[q(0)-a_q]^2/2}$ while the bath still resides in the 
state~(\ref{eq cond eq bath}). The fact that the bath has no time to adapt to a new conditional equilibrium state causes 
non-Markovian behaviour as we can rigorously show with our quantifier. For instance, 
in view of Eq.~(\ref{eq time homo Fourier}) we find that 
\begin{equation}
 \begin{split}
  & -i\frac{d}{d\omega}\tilde\chi(\omega) - \tilde\chi(\omega)\chi_+^{-1}\tilde\chi(\omega) 	\\
  & = \frac{D}{(\omega_0^2 - \omega^2 - iD\omega)^2}
    \left(\begin{array}{cc}
	   1		& i\omega	\\
	   -i\omega	& \omega^2	\\
	  \end{array}\right),
 \end{split}
\end{equation}
which is clearly non-zero and only becomes negligible in the weak coupling regime, see Fig.~\ref{fig Ohmic}. 
The subtle importance of the initial state preparation procedure for the validity of the Langevin equation was already noted in 
Ref.~\cite{BezZPB1980, SanchezCanizaresSolsPhysA1994, HaenggiLectNotes1997}, but it had not been rigorously quantified. 

We remark that it is a special property of the Caldeira-Leggett model that the first moments do not depend on 
$\hbar$. This changes for CFs, which depend on $\hbar$ and have a non-trivial dependence on the inverse bath temperature, 
see Fig.~\ref{fig Ohmic} again. 

\begin{figure*}%[b]
 \centering\includegraphics[width=0.99\textwidth,clip=true]{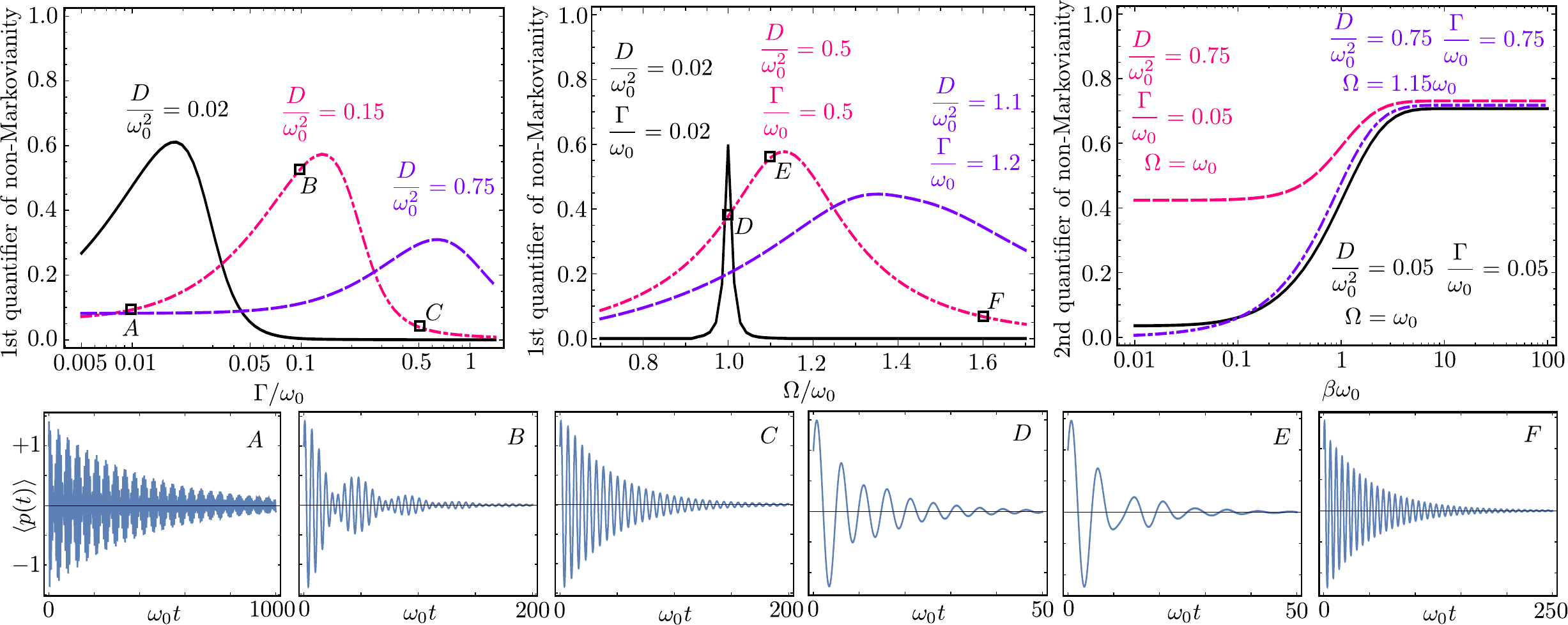}
 \label{fig plot peaked SD} 
 \caption{\bb{Top row:} We use the convention of Fig.~\ref{fig Ohmic} where a solid line refers to $\C N_{qq}^{(1/2)}$, a 
 dashed line to $\C N_{qp}^{(1/2)}$ and a dash-dotted line to $\C N_{pp}^{(1/2)}$, but the colour coding is different. 
 \emph{Left:} Plot (in logarithmic scale) over the dimensionless parameter $\Gamma/\omega_0$, which controls the sharpness 
 of the peak, at resonance ($\Omega = \omega_0$) for increasing coupling strengths (following the position of 
 the peaks from left to right) $D = 0.02\omega_0^2$ (black solid line), $D = 0.15\omega_0^2$ (pink dash-dotted line) and 
 $D = 0.75\omega_0^2$ (purple dashed line). \emph{Middle:} Plot over the dimensionless resonance frequency 
 $\Omega/\omega_0$ of the bath for increasing $(D,\Gamma)$ (following the positions of the peak from left to right) 
 $D = 0.05\omega_0^2, \Gamma = 0.05\omega_0$ (black solid line), $D = 0.25\omega_0^2, \Gamma = 0.25\omega_0$ (pink 
 dash-dotted line) and $D = 1.1\omega_0^2, \Gamma = 1.2\omega_0$ (purple dashed line). \emph{Right:} Plot over the 
 dimensionless inverse temperature $\beta\omega_0$ in logarithmic scale for 
 $D = 0.05\omega_0^2, \Gamma = 0.05\omega_0, \Omega = \omega_0$ (black solid line at bottom), 
 $D = 0.75\omega_0^2, \Gamma = 0.05\omega_0, \Omega = \omega_0$ (pink dashed line on top) and 
 $D = 0.75\omega_0^2, \Gamma = 0.75\omega_0, \Omega = 1.15\omega_0$ (purple dash-dotted line). 
 We set $\hbar\equiv 1$. \bb{Bottom row:} The time evolution of the system momentum $\lr{p(t)}$ is exemplarily 
 depicted for six different parameter values indicated by the letters $A,B,C,D,E$ and $F$ and a solid black square 
 in the plots of the top row. The initial perturbation was choosen to be $a_p/\sqrt{\omega_0} = \sqrt{\omega_0} a_q = 1$. }
\end{figure*}

{\it Peaked SD.---} We now turn to a non-trivial case described by the SD 
\begin{equation}
 J(\omega) = \frac{D^2\Gamma\omega}{(\omega^2-\Omega^2)^2 + \Gamma^2\omega^2}. 
\end{equation}
This corresponds to the SD felt by a system, which is coupled with strength $D$ to another harmonic oscillator of 
frequency $\Omega$, which is in turn coupled to an Ohmic bath with SD $\Gamma\omega$~\cite{GargOnuchicAmbegaokarJCP1985}. 
Note that the parameter $\Gamma$ controls the structure of the SD: a small $\Gamma$ corresponds to a sharp peak around 
the frequency $\Omega$ whereas a larger $\Gamma$ smears out the peak resulting in an increasingly flat SD. 
Furthermore, the real part of the Fourier transformed memory kernel is $\Re[\tilde\gamma(\omega)] = J(\omega)/\omega$ 
and the imaginary part becomes (see supplementary material for more details) 
\begin{equation}\label{eq peaked SD im gamma}
 \Im[\tilde\gamma(\omega)] = \frac{\Gamma^2 + \omega^2 - \Omega^2}{\Gamma\Omega^2} J(\omega).
\end{equation}

In practical considerations, non-Markovian behaviour is often associated with a strong system-bath coupling and a 
structured SD~\cite{DeVegaAlonsoRMP2017}. Thus, one would intuitively expect that the degree of non-Markovianity increases 
for larger $D$ and smaller $\Gamma$ and that it reaches a maximum, if the system is on resonance with the oscillator 
in the bath, i.e., if $\omega_0\approx\Omega$. As Fig.~\ref{fig plot peaked SD} demonstrates, this intuition is not 
always correct. We observe that there is no simple (i.e., monotonic) relation between our quantifier of 
non-Markovianity and the parameters $D$, $\Gamma$ and $|\omega_0 - \Omega|$. 
In fact, one could ask whether this results from the particular definition~(\ref{eq distance f g}) 
and~(\ref{eq quantifier 1}) which we have used and which always entails a certain level of arbitrariness. Therefore, we 
have also plotted the time-evolution of the observable $\lr{p(t)}$ in Fig.~\ref{fig plot peaked SD} (bottom row), whose 
deviation from an exponentially damped oscillation seems to be roughly in agreement with our quantification scheme.

{\it Summary.---} This work shows that it is possible to quantify non-Markovianity in the linear response regime in a 
rigorous and straightforward manner. Since we can only treat initially correlated states, our approach is rather 
``orthogonal'' to previous ones, but for many scenarios of experimental interest 
this might be indeed a more realisitic assumption. Furthermore, for the Caldeira-Leggett 
model our quantifier is valid beyond linear response and can be expressed analytically in terms of an integral over known 
functions. We have then shown that even a classical particle coupled to an Ohmic bath can behave non-Markovian depending on 
the initial state preparation procedure and that one should not expect a simple relation between our quantifier of 
non-Markovianity and parameters in the SD or the temperature of the bath. 

{\it Acknowledgements.---} We are grateful to Victor Bastidas, John Bechhoefer and Javier Cerrillo for discussions and 
comments. This research is funded by the European Research Council project NanoThermo (ERC-2015-CoG Agreement 
No. 681456).

%%%%%%%%%%%%%%%%%%%%%%%%%%%%%%%%%%%%%%%%%%%%%%%%%%%%%%%%%%%%%%%%%%%%%%%%%%%%%%%%%%%%%%%%%%%%%%%%%%%%%%%%%%%%%%%%%%%%%%%%

\bibliography{/home/philipp/Documents/references/books,/home/philipp/Documents/references/general_refs,/home/philipp/Documents/references/open_systems,/home/philipp/Documents/references/thermo,/home/philipp/Documents/references/info_thermo,/home/philipp/Documents/references/general_QM}

%%%%%%%%%%%%%%%%%%%%%%%%%%%%%%%%%%%%%%%%%%%%%%%%%%%%%%%%%%%%%%%%%%%%%%%%%%%%%%%%%%%%%%%%%%%%%%%%%%%%%%%%%%%%%%%%%%%%%%%%
\appendix

%%%%%%%%%%%%%%%%%%%%%%%%%%%%%%%%%%%%%%%%%%%%%%%%%%%%%%%%%%%%%%%%%%%%%%%%%%%%%%%%%%%%%%%%%%%%%%%%%%%%%%%%%%%%%%%%%%%%%%%%
\section{SUPPLEMENTARY MATERIAL}

This appendix contains in the following order: 
\begin{itemize}
 \item A detailed discussion about the existence of the mean value propagator $G(t)$.
 \item A derivation of Eq.~(\ref{eq time homo Fourier}). 
 \item A detailed discussion and derivation of the RT~(\ref{eq RT}). 
 \item A derivation of Eq.~(\ref{eq CF prediction RT general}). 
 \item A derivation ot the matrix of linear response functions~(\ref{eq response fct Brownian}). 
 \item A derivation of the generalized Langevin equation~(\ref{eq Langevin anormal}). 
 \item A derivation of a general relation between the Fourier transformed memory kernel and the spectral density 
 as well as a derivation of Eq.~(\ref{eq peaked SD im gamma}). 
\end{itemize}
For notational convenience, we will often denote 
\begin{equation}
 \chi_+^{-1} = \frac{1}{\chi_+}.
\end{equation}

%%%%%%%%%%%%%%%%%%%%%%%%%%%%%%%%%%%%%%%%%%%%%%%%%%%%%%%%%%%%%%%%%%%%%%%%%%%%%%%%%%%%%%%%%%%%%%%%%%%%%%%%%%%%%%%%%%%%%%%%
\section{Existence of $G(t)$}

The existence of $G(t)$ is guaranteed if the matrix $\chi_+$ can be inverted. To approach the problem, we first consider 
the simplest case where the perturbation is caused only by a single observable $A_1$. Then, $\chi_+$ is the scalar 
\begin{equation}
 \chi_+ = \frac{i}{\hbar}\lr{[A_1,A_1]}_\beta = 0.
\end{equation}
Clearly, in this case $\chi_+$ can never be inverted. 

To study the Caldeira-Leggett model in the main text, a perturbation caused by two observables $A_1$ and $A_2$ turns 
out to be sufficient. In fact, one can in general hope that this is always the case because 
\begin{equation}
 \chi_+ = \frac{i}{\hbar}\lr{[A_1,A_2]}_\beta 
 \left(\begin{array}{cc}
        0	&	1	\\
        -1	&	0	\\
       \end{array}\right).
\end{equation}
Hence, if we choose $A_1$ and $A_2$ such that $\lr{[A_1,A_2]}_\beta \neq 0$, the existence of the inverse of 
$\chi_+$ is guaranteed. As an additional example we consider an open two-level system (TLS) such as the spin-boson model, 
but the particular form of the environment is unimportant for the present reasoning. Let us choose a basis such that 
the reduced equilibrium state of the TLS is aligned along the Pauli-matrix $\sigma_z$ such that 
$\lr{\sigma_z}_\beta \neq 0$. Then, if we decide to perturb the system using the observables $A_1 = \sigma_x$ and 
$A_2 = \sigma_y$, we obtain 
\begin{equation}
 \chi_+ = \frac{2i}{\hbar}
 \left(\begin{array}{cc}
        0	&	\lr{\sigma_z}_\beta	\\
        -\lr{\sigma_z}_\beta 	&	0	\\
       \end{array}\right),
\end{equation}
which is invertible. 

Finally, it is interesting to ask what happens if we choose a basis set of $N^2-1$ observables in case of an 
$N$-dimensional quantum system. As the expectation values of these observables is isomorphic to the complete system 
density operator, this would provide us with a full dynamical map for the system and questions related to its 
complete positivity would then be relevant. Let us start with the TLS from before and let us add $A_3 = \sigma_z$ to the 
set of observables $\{A_1,A_2\}$. Then, 
\begin{equation}
 \chi_+ = \frac{2i}{\hbar}
 \left(\begin{array}{ccc}
        0			&	\lr{\sigma_z}_\beta	&	-\lr{\sigma_y}_\beta	\\
        -\lr{\sigma_z}_\beta 	&	0			&	\lr{\sigma_x}_\beta	\\
        \lr{\sigma_y}_\beta	&	-\lr{\sigma_x}_\beta	&	0			\\
       \end{array}\right).
\end{equation}
It is easy to confirm that the determinant of this matrix is zero, and hence $\chi_+$ is not invertible. 
Indeed, this will always be the case. To demonstrate this we show that there are in case of a complete set 
of $N^2-1$ observables multiple $\bb a$'s, which give rise to the same state of the system and hence, $\chi_+$ maps 
many to one and is not invertible. For this purpose we introduce the effective Hamiltonian $H_\text{eff}$ 
which describes the reduced Gibbs state of the system prior to the perturbation,
\begin{equation}
 \rho_S(t_0) = \mbox{tr}_B\{e^{-\beta(H_S+H_I+H_B)}\}/Z \equiv e^{-\beta H_\text{eff}},
\end{equation}
where $Z$ is the global partition function. As the set $A_i$ is assumed to be a complete set of observables, 
we can actually expand the effective Hamiltonian, too, 
\begin{equation}
 H_\text{eff} = b_0 1_S + \sum_{i=1}^N b_i A_i
\end{equation}
for some real-valued parameters $b_0, b_1,\dots b_N$. Let us now choose 
$\bb a = \epsilon(b_1,\dots,b_N)$ where $\epsilon$ is some constant choosen small enough such that $\bb a$ can 
be safely treated within linear response. Then, the net effect will be that $U_0$ and $H_\text{eff}$ commute such that we 
obtain the same initial state 
$\rho_S(t_0) = \mbox{tr}_B\{U_0 e^{-\beta(H_S+H_I+H_B)}U_0^\dagger\}/Z = e^{-\beta H_\text{eff}}$ independent of the 
choice of $\epsilon$. Since there are multiple $\epsilon$ possible, this proves that $\chi_+$ can not be invertible.

%%%%%%%%%%%%%%%%%%%%%%%%%%%%%%%%%%%%%%%%%%%%%%%%%%%%%%%%%%%%%%%%%%%%%%%%%%%%%%%%%%%%%%%%%%%%%%%%%%%%%%%%%%%%%%%%%%%%%%%%
\section{Derivation of Eq.~(\ref{eq time homo Fourier})}

The assumed divisibility property~(\ref{eq time homogeneity}) depends explicitly on the choice of $s\in[0,t]$. 
To get a simple and average quantifier, we decide to integrate Eq.~(\ref{eq time homogeneity}) over $s$ 
from zero to $t$. This yields 
\begin{equation}
 \int_0^t ds \chi(t) = t\chi(t) = \int_0^t ds \chi(t-s)\frac{1}{\chi_+}\chi(s),
\end{equation}
Now, we recognize that due to the Heaviside step functions involved in the definition of $\chi(t)$, we can write the 
right hand side as 
\begin{equation}
 \int_0^t ds \chi(t-s)\frac{1}{\chi_+}\chi(s) = \int_{-\infty}^\infty ds \chi(t-s)\frac{1}{\chi_+}\chi(s). 
\end{equation}
Fourier transformation then immediately yields 
\begin{equation}
 -i\frac{d}{d\omega}\tilde\chi(\omega) = \tilde\chi(\omega)\frac{1}{\chi_+}\tilde\chi(\omega),
\end{equation}
as claimed in the main text.

%%%%%%%%%%%%%%%%%%%%%%%%%%%%%%%%%%%%%%%%%%%%%%%%%%%%%%%%%%%%%%%%%%%%%%%%%%%%%%%%%%%%%%%%%%%%%%%%%%%%%%%%%%%%%%%%%%%%%%%%
\section{Derivation of the regression theorem}

We here state, prove and discuss the quantum regression theorem following a rather general approach. The discussion of 
the RT in the current literature is in fact often motiviated by a particular physical situation. \\

\textbf{Theorem.} 
 {\it If there exists a set of system observables $\{A_i\}$ such that their dynamics is closed, i.e., if }
 \begin{equation}\label{eq assump 1 RT}
  \lr{A_i(t)} = \sum_j G_{ij}(t,t_0)\lr{A_j(t_0)}
 \end{equation}
 {\it for all times $t\ge t_0$ and for some propagator $G(t,t_0)$ independent of the initial state $\rho_S(t_0)$, 
 and if there exists a dynamical map $\Phi(t,t_0)$ independent of the initial state $\rho_S(t_0)$, then }
 \begin{equation}\label{eq RT app}
  \lr{A_i(t)A_j(t_0)} = \sum_k G_{ik}(t,t_0)\lr{A_k(t_0)A_j(t_0)}
 \end{equation}
 {\it for all times $t\ge t_0$.} \\
 
\textbf{Proof.}
 According to quantum mechanics, the expectation value of a system observable is given 
 by\footnote{We are particularly cautious here and also indicate the ``time-dependence'' on $t_0$ for an operator in 
 the Schr\"odinger picture. Clearly, $t_0$ is the time choosen were the Schr\"odinger and Heisenberg picture coincide. } 
 \begin{equation}\label{eq proof RT 1}
  \begin{split}
   \lr{A_i(t)}	&=	\mbox{tr}_S\{A_i(t_0)\rho_S(t)\}	\\
		&=	\mbox{tr}_S\{A_i(t_0)\Phi(t,t_0)\rho_S(t_0)\}.
  \end{split}
 \end{equation}
 We add that the existence of some CPTP map $\Phi(t,t_0)$, which maps the initial state $\rho_S(t_0)$ to the final state 
 $\rho_S(t)$, is always guaranteed~\cite{WuEtAlJPA2007}, but in general its construction is highly non-unique and 
 depends on $\rho_S(t_0)$. 
 
 We now use our two assumptions. Namely, by comparing Eq.~(\ref{eq proof RT 1}) with Eq.~(\ref{eq assump 1 RT}) and by 
 using that $\Phi(t,t_0)$ is the correct dynamical map for any initial system state, we conclude that 
 \begin{equation}
  \begin{split}\label{eq help}
   & \mbox{tr}_S\left\{A_i(t_0)\Phi(t,t_0)\rho_S(t_0)\right\}	\\
   & = \sum_j G_{ij}(t,t_0) \mbox{tr}_S\left\{A_j(t_0)\rho_S(t_0)\right\}
  \end{split}
 \end{equation}
 must hold for any $\rho_S(t_0)$. By using that 
 \begin{equation}
  \begin{split}
   & \mbox{tr}_S\{[\Phi^*(t,t_0)A_i(t_0)]A_j(t_0)\rho_S(t_0)\}	\\
   & = \mbox{tr}_S\left\{A_i(t_0)\Phi(t,t_0)[A_j(t_0)\rho_S(t_0)]\right\},
  \end{split}
 \end{equation}
 where $\Phi^*(t,t_0)$ is the adjoint dynamical map in the Heisenberg picture, we obtain with the help of 
 Eq.~(\ref{eq help}) the chain of equalities 
 \begin{equation}
  \begin{split}
   \lr{A_i(t)A_j(t_0)}	&=	\mbox{tr}_S\left\{A_i(t_0)\Phi(t,t_0)[A_j(t_0)\rho_S(t_0)]\right\}	\\
			&=	\sum_k G_{ik}(t,t_0)\mbox{tr}_S\left\{A_k(t_0) A_j(t_0)\rho_S(t_0)\right\}	\\
			&=	\sum_k G_{ik}(t,t_0)\lr{A_k(t_0) A_j(t_0)},	\nonumber
  \end{split}
 \end{equation}
 which is the RT~(\ref{eq RT app}). QED. \\

\textbf{Remarks.} From the proof above it becomes evident that we neither had to use the quantum Chapman-Kolmogorov 
equation~(\ref{eq CP divisibility}) nor any product state assumption for the system-bath state. Physically, of course, 
we know that deriving a dynamical map $\Phi(t,t_0)$ is often admissible only for factorized initial state. However, 
also purely classically correlated initial states yield to a CPTP map $\Phi(t,t_0)$~\cite{RodriguezRosarioEtAlJPA2008} 
and even in presence of quantum correlations it is sometimes possible to derive such maps~\cite{BroducthEtAlPRA2013, 
BuscemiPRL2014, DominyShabaniLidarQIP2016}. Therefore, it is important to distinguish the question {\it When can we 
derive $\Phi(t,t_0)$ in a physical setting?} from {\it What does an assumed existence of such a $\Phi(t,t_0)$ imply?} 

Within our context the validity of the RT and the value of the corresponding quantifier $\C N_{ij}^{(2)}$ in 
Eq.~(\ref{eq quantifier 2}) give us key information about the question whether knowledge of the initial system state 
and its evolution suffices to infer the evolution of the (equilibrium) CFs, but it does not reveal any more insights. 

Moreover, as Figs.~\ref{fig Ohmic} and~\ref{fig plot peaked SD} explicitly demonstrate, the RT can even fail in the 
classical limit $\beta\rightarrow0$. The reason for that can be traced back to the fact that the propagator $G(t)$ in 
Eq.~(\ref{eq Kubo}) is only well-defined for $a_i\neq 0$ and cannot be used at equilibrium. Thus, our first assumption 
that $G(t)$ is the correct propagator for the mean values for any initial system state $\rho_S(t_0)$ was clearly wrong.

%%%%%%%%%%%%%%%%%%%%%%%%%%%%%%%%%%%%%%%%%%%%%%%%%%%%%%%%%%%%%%%%%%%%%%%%%%%%%%%%%%%%%%%%%%%%%%%%%%%%%%%%%%%%%%%%%%%%%%%%
\section{Equilibrium correlation functions from the regression theorem}

At equilibrium we have the symmetry $\lr{A_i(-t)A_j}_\beta = \lr{A_iA_j(t)}_\beta$ and using the RT we obtain 
$\C C^\text{RT}(-t) = \C C(0)^T G(t)^T$ for $t>0$ where $T$ denotes the transpose and $\C C(0)$ the initial equilibrium 
covariance matrix. In terms of the linear response function, we can also write 
$\C C^\text{RT}(-t) = -\C C(0)^T\chi_+^{-1}\chi(t)^T$ where we used the skew symmetry $\chi_+ = -\chi_+^T$. 
Therefore, Fourier transformation yields 
\begin{equation}
 \begin{split}
  & \tilde{\C C}^\text{RT}(\omega)	\\
  & = \int_0^\infty dt e^{i\omega t} \chi(t)\frac{1}{\chi_+}\C C(0) - \int_{-\infty}^0 dt e^{i\omega t}\C C(0)^T\frac{1}{\chi_+}\chi(-t)^T	\\
  & = \int_0^\infty dt e^{i\omega t} \chi(t)\frac{1}{\chi_+}\C C(0) - \int_0^\infty dt e^{-i\omega t}\C C(0)^T\frac{1}{\chi_+}\chi(t)^T	\\
  &= \tilde\chi(\omega)\frac{1}{\chi_+}\C C(0) - \C C(0)\frac{1}{\chi_+}\tilde\chi(-\omega)^T.
 \end{split}
\end{equation}
For Hermitian observables it now holds true that $\tilde\chi(-\omega) = \tilde\chi^*(\omega)$ such that we obtain 
\begin{equation}
 \tilde{\C C}^\text{RT}(\omega) = \tilde\chi(\omega)\frac{1}{\chi_+}\C C(0) - \C C(0)\frac{1}{\chi_+}\tilde\chi(\omega)^\dagger
\end{equation}
as claimed in the main text. More explicitly, using Eq.~(\ref{eq response fct Brownian}), we get 
\begin{align}
 \tilde{\C C}^\text{RT}_{qq}(\omega)	&=	2\omega C_{qq}(0)\Im[\tilde\chi_{qq}],	\\
 \tilde{\C C}^\text{RT}_{qp}(\omega)	&=	C_{pp}(0)\tilde\chi_{qq} - C_{qq}(0)(\omega^2 \tilde\chi^*_{qq}+1),	\\
 \tilde{\C C}^\text{RT}_{pq}(\omega)	&=	C_{pp}(0)\tilde\chi^*_{qq} - C_{qq}(0)(\omega^2 \tilde\chi_{qq}+1),	\\
 \tilde{\C C}^\text{RT}_{pp}(\omega)	&=	2\omega C_{pp}(0)\Im[\tilde\chi_{qq}].
\end{align}
Note the symmetry relation $\tilde{\C C}_{qp}^\text{RT}(\omega) = \tilde{\C C}_{pq}^\text{RT}(\omega)^*$. 
This has to be compared with the true CF, which follows from Eq.~(\ref{eq relation eq CFs}) and the fluctuation 
dissipation theorem~(\ref{eq correlation FDT}): 
\begin{equation}
 \tilde{\C C}(\omega) = \frac{2\hbar \Im[\chi_{qq}(\omega)]}{1-e^{-\beta\hbar\omega}} 
 \left(\begin{array}{cc}
        1		&	i\omega	\\
        -i\omega	&	\omega^2	\\
       \end{array}\right).
\end{equation}

%%%%%%%%%%%%%%%%%%%%%%%%%%%%%%%%%%%%%%%%%%%%%%%%%%%%%%%%%%%%%%%%%%%%%%%%%%%%%%%%%%%%%%%%%%%%%%%%%%%%%%%%%%%%%%%%%%%%%%%%
\section{Response function of the Brownian motion model}

We here derive Eq.~(\ref{eq response fct Brownian}) using the well-known result for the position-position response 
function~\cite{GrabertSchrammIngoldPR1988, WeissBook2008} 
\begin{equation}
 \tilde\chi_{qq}(\omega) = \frac{1}{\omega_0^2 - \omega_2 - i\omega\tilde\gamma(\omega)}.
\end{equation}

To derive the other response functions, one uses (i) that 
\begin{equation}
 \frac{d}{dt}q(t) = \frac{i}{\hbar}[H(t),q(t)] = p(t),
\end{equation}
where $H$ denotes the unperturbed Hamiltonian~(\ref{eq Hamiltonian}) of the Caldeira-Leggett model, (ii) partial 
integration and (iii) $\lr{A_i(t)A_j(0)}_\beta = \lr{A_i(0)A_j(-t)}_\beta$ where we explicitly wrote out the 
``dependence'' of the operators on the initial time $t_0 = 0$ in the Heisenberg picture. 

Then, in detail the other response functions can be derived as follows: 
\begin{equation}
 \begin{split}
  \tilde\chi_{qp}(\omega)	&=	\frac{i}{\hbar}\int_0^\infty dt e^{i\omega t}\lr{[q(t),p(0)]}_\beta	\\
				&=	\frac{i}{\hbar}\int_0^\infty dt e^{i\omega t}\lr{[q(0),p(-t)]}_\beta	\\
				&=	\frac{i}{\hbar}\int_0^\infty dt e^{i\omega t}\left(-\frac{d}{dt}\lr{[q(t),q(0)]}_\beta\right)	\\
				&=	-\frac{i}{\hbar}\left.e^{i\omega t}\lr{[q(t),q(0)]}_\beta\right|_0^\infty + i\omega \tilde\chi_{qq}(\omega)	\\
				&=	i\omega \tilde\chi_{qq}(\omega),
 \end{split}
\end{equation}
where we used $\lim_{t\rightarrow\infty}\lr{[q(t),q(0)]}_\beta = 0$ and $[q(0),q(0)] = 0$ at the end. 
Analogously, one also derives $\tilde\chi_{pq}(\omega) = -i\omega \tilde\chi_{qq}(\omega)$. To derive the 
corresponding result for $\tilde\chi_{pp}(\omega)$, two partial integrations are necessary: 
\begin{equation}
 \begin{split}
  \tilde\chi_{pp}(\omega)	&=	\frac{i}{\hbar}\int_0^\infty dt e^{i\omega t}\frac{d}{dt}\lr{[q(t),p(0)]}_\beta	\\
				&=	\frac{i}{\hbar}\left.e^{i\omega t}\lr{[q(t),p(0)]}_\beta\right|_0^\infty - i\omega \tilde\chi_{qp}(\omega)	\\
				&=	-\frac{i}{\hbar} \lr{i\hbar}_\beta - i\omega [i\omega\tilde\chi_{qq}(\omega)]	\\
				&=	1 + \omega^2\tilde\chi_{qq}(\omega).
 \end{split}
\end{equation}

%%%%%%%%%%%%%%%%%%%%%%%%%%%%%%%%%%%%%%%%%%%%%%%%%%%%%%%%%%%%%%%%%%%%%%%%%%%%%%%%%%%%%%%%%%%%%%%%%%%%%%%%%%%%%%%%%%%%%%%%
\section{Generalized Langevin equation}

We here go through the derivation of the generalized Langevin equation for the Caldeira-Leggett 
model~(\ref{eq Hamiltonian}) with an arbitrary linear system force added to it, 
\begin{equation}
 H(t) = H - f_q(t)q - f_p(t)p.
\end{equation}
Our treatment is only slightly more general than the one in Ref.~\cite{WeissBook2008}, where a momentum-dependent 
force is not considered. Below we will assume that 
right before $t_0 = 0$ the system and the bath is in a global equilibrium state $\omega(\beta)$ and we 
explicitly require that the perturbation is switched on at times $t\ge0$, i.e., $f_q(t) = 0 = f_p(t)$ for $t<0$. 

The equations of motion for the positions and momenta of the system and the bath become (note that there form is 
identical classically and quantum mechanically) 
\begin{align}
 \dot q(t)	&=	p(t) - f_p(t),	\nonumber	\\
 \dot p(t)	&=	-\omega_0^2 q(t) - \sum_k \frac{c_k^2}{\omega_k^2}q(t) + \sum_k c_k q_k(t) + f_q(t),	\nonumber	\\
 \dot q_k(t)	&=	p_k(t),	\nonumber	\\
 \dot p_k(t)	&=	-\omega_k^2 q_k(t) + c_k q(t).	\nonumber
\end{align}
The last two equations are formally solved by 
\begin{equation}
 \begin{split}
  q_k(t)	=&~	q_k(0) \cos(\omega_k t) + \frac{p_k(0)}{\omega_k}\sin(\omega_k t)	\\
		&+	\frac{c_k}{\omega_k} \int_{0}^t ds \sin[\omega_k(t-s)]q(s).	\nonumber
 \end{split}
\end{equation}
It turns out to be convenient to rewrite the integral using partial integration: 
\begin{equation}
 \begin{split}
  & \int_{0}^t ds \sin[\omega_k(t-s)]q(s)	\\
  & = \frac{q(t)}{\omega_k} - \frac{q(0)}{\omega_k}\cos(\omega_k t) - \frac{1}{\omega_k}\int_{0}^t ds \cos[\omega_k(t-s)]\dot q(s).	\nonumber
 \end{split}
\end{equation}
Note that in presence of $f_p(t)$ it is in general not valid to replace $\dot q$ by $p$ under the integral. 
With the help of this formal solution we obtain 
\begin{align}
 \dot q(t)	=&~	p(t) - f_p(t),	\nonumber	\\
 \dot p(t)	=&	-\omega_0^2 q(t) + f_q(t) + \sum_k c_k q_k(0) \cos(\omega_k t)	\nonumber	\\
		&+	\sum_k \frac{c_k p_k(0)}{\omega_k}\sin(\omega_k t) - \sum_k \frac{c_k^2}{\omega_k^2} q(0)\cos(\omega_k t)	\nonumber	\\
		&-	\sum_k \frac{c_k^2}{\omega_k^2} \int_{0}^t ds \cos[\omega_k(t-s)]\dot q(s).
\end{align}
We define the noise 
\begin{equation}
 \begin{split}
  \xi(t)	=&~	\sum_k c_k q_k(0) \cos(\omega_k t) + \sum_k \frac{c_k p_k(0)}{\omega_k}\sin(\omega_k t)	\nonumber	\\
		&-	\sum_k \frac{c_k^2}{\omega_k^2} q(0)\cos(\omega_k t),	\nonumber
 \end{split}
\end{equation}
which obeys in the classical case $\lr{\lr{\xi(t)}} = 0$ and $\lr{\lr{\xi(t)\xi(s)}} = \gamma(t-s)/\beta$, where the 
double bra-ket notation refers to an average over the conditionally equilibrated bath state in Eq.~(\ref{eq cond eq bath}). 
Using also the definition of the SD and the memory kernel, we can write 
\begin{equation}
 \begin{split}
  & \sum_k \frac{c_k^2}{\omega_k^2} \int_{0}^t ds \cos[\omega_k(t-s)]\dot q(s)	\\
  & = \frac{2}{\pi}\int_0^\infty d\omega\frac{J(\omega)}{\omega}\int_{0}^t ds \cos[\cos(\omega(t-s)]\dot q(s)	\\
  & = \int_{0}^\infty ds \gamma(t-s)\dot q(s),
 \end{split}
\end{equation}
which allows us to arrive at the more compact expression 
\begin{align}
 \dot q(t)	&=	p(t) - f_p(t),	\label{eq Langevin general}	\\
 \dot p(t)	&=	-\omega_0^2 q(t) + f_q(t) - \int_{0}^\infty ds \gamma(t-s)\dot q(s) + \xi(t).	\nonumber
\end{align}
This equation generalized  Langevin equation reduces to Eq.~(\ref{eq Langevin anormal}) in the limit of an Ohmic SD.

%%%%%%%%%%%%%%%%%%%%%%%%%%%%%%%%%%%%%%%%%%%%%%%%%%%%%%%%%%%%%%%%%%%%%%%%%%%%%%%%%%%%%%%%%%%%%%%%%%%%%%%%%%%%%%%%%%%%%%%%
\section{Fourier transformed memory kernel}

For the analysis of our non-Markovianity witness it is important to be able to compute the Fourier transform 
$\tilde\gamma(\omega)$ of the memory kernel 
$\gamma(t) = \Theta(t)\frac{2}{\pi}\int_0^\infty d\omega \frac{J(\omega)}{\omega}\cos(\omega t)$, which enters the 
generalized Langevin equation. We start with 
\begin{equation}
 \tilde\gamma(\omega) = \int_0^\infty d\nu \frac{J(\nu)}{\pi\nu} \int_0^\infty dt [e^{i(\omega+\nu)t} + e^{i\omega-\nu)t}]
\end{equation}
and use the theorem 
\begin{equation}
 \int_0^\infty dt e^{\pm i\omega t} = \pi\delta(\omega) \pm i\C P\frac{1}{\omega} = \pm i\lim_{\epsilon\searrow0} \frac{1}{x\pm i\epsilon}.
\end{equation}
This yields 
\begin{equation}
 \tilde\gamma(\omega) = \lim_{\epsilon\searrow0} \int_0^\infty d\nu \frac{J(\nu)}{\pi\nu} \left(\frac{i}{\nu+\omega+i\epsilon} - \frac{i}{\nu-\omega-i\epsilon}\right).
\end{equation}
Splitting everything into real and imaginary part and using the identity 
\begin{equation}
 \pi\delta(x) = \lim_{\epsilon\searrow0}\frac{\epsilon}{x^2+\epsilon^2},
\end{equation}
we deduce that 
\begin{align}
 \Re[\tilde\gamma(\omega)]	=&~	\frac{J(\omega)}{\omega},	\\
 \Im[\tilde\gamma(\omega)]	=&~	\lim_{\epsilon\searrow0} \int_0^\infty d\nu \frac{J(\nu)}{\pi\nu} \frac{\nu+\omega}{(\nu+\omega)^2+\epsilon^2}	\\
				&-	\lim_{\epsilon\searrow0} \int_0^\infty d\nu \frac{J(\nu)}{\pi\nu} \frac{\nu-\omega}{(\nu-\omega)^2+\epsilon^2}	\nonumber.
\end{align}
If we use the the standard convention $J(-\omega) = J(-\omega)$ to extend the SD to negative frequencies, 
the imaginary part can be also expressed as 
\begin{equation}
 \begin{split}\label{eq memory kernel im}
  \Im[\tilde\gamma(\omega)]	=&~	\lim_{\epsilon\searrow0} \int_{-\infty}^\infty d\nu \frac{J(\nu)}{2\pi\nu} \frac{\nu+\omega}{(\nu+\omega)^2+\epsilon^2}	\\
				&-	\lim_{\epsilon\searrow0} \int_{-\infty}^\infty d\nu \frac{J(\nu)}{2\pi\nu} \frac{\nu-\omega}{(\nu-\omega)^2+\epsilon^2},
 \end{split}
\end{equation}
which is advantageous for the peaked SD considered in the main text. 

Let us abbreviate Eq.~(\ref{eq memory kernel im}) as 
\begin{equation}
 \Im[\tilde\gamma(\omega)] = \lim_{\epsilon\searrow0}\left\{\int_{-\infty}^\infty d\nu f(\nu,\omega) - \int_{-\infty}^\infty d\nu f(\nu,-\omega)\right\}
\end{equation}
with 
\begin{equation}
 f(\nu,\omega) = \frac{1}{2\pi}\frac{D^2\Gamma}{(\omega^2-\Omega^2)^2 + \Gamma^2\omega^2} \frac{\nu+\omega}{(\nu+\omega)^2+\epsilon^2}.
\end{equation}
To evaluate the integral we use the residue theorem in the upper half complex plane. For this purpose we note that the 
roots of $f(\nu,\omega)$ are given by 
\begin{align}
 r_1	&=	\sqrt{\frac{c-\Gamma^2}{4} + i\frac{\Gamma\sqrt{c}}{2}}, ~ r_4 = \sqrt{\frac{c-\Gamma^2}{4} - i\frac{\gamma\sqrt{c}}{2}},	\nonumber	\\
 r_2	&=	-\sqrt{\frac{c-\Gamma^2}{4} - i\frac{\Gamma\sqrt{c}}{2}}, ~ r_5 = -\sqrt{\frac{c-\Gamma^2}{4} + i\frac{\gamma\sqrt{c}}{2}},	\nonumber	\\
 r_3	&=	-\omega+i\epsilon, ~ r_6 = -\omega-i\epsilon.	\nonumber
\end{align}
Here, we defined $c \equiv 4\Omega^2 - \Gamma^2$ and we assume that $2\Omega^2 - \Gamma^2 > 0$ in the following. 
Then, by defining the square root of a complex number $z = re^{i\phi}$ in the canonical way, 
$\sqrt{z} = \sqrt{r}e^{i\phi/2}$, we recognize that the first three roots lie in the upper half complex plane. 
Thus, since we have only simple first order poles, the residue theorem yields 
\begin{equation}
 \int_{-\infty}^\infty d\nu f(\nu,\omega) = 2\pi i\sum_{i=1}^3 \lim_{\nu\rightarrow r_i}(\nu-r_i)f(\nu,\omega).
\end{equation}
Explicit evaluation of the two integrals, where the second can be obtained from the first by mapping 
$\omega\mapsto-\omega$, gives 
\begin{equation}
 \begin{split}
  & \int_{-\infty}^\infty d\nu f(\nu,\omega) - \int_{-\infty}^\infty d\nu f(\nu,-\omega) = 	\\
  & \frac{16 D^2 \omega[3\Gamma^2 + 8\Gamma\epsilon - c + 4(\omega^2+\epsilon^2)]}{(\Gamma^2+c) [(\sqrt{c} - 2\omega)^2 + (\Gamma + 2\epsilon)^2] [(\sqrt{c} + 2\omega)^2 + (\Gamma + 2\epsilon)^2]}
 \end{split}
\end{equation}
It is straightforward to take the limit $\epsilon\searrow0$, which yields 
\begin{equation}
 \begin{split}
  \Im[\tilde\gamma(\omega)]	&=	\frac{16 D^2 \omega[3\Gamma^2 - c + 4\omega^2]}{(\Gamma^2+c) [(\sqrt{c} - 2\omega)^2 + \Gamma^2] [(\sqrt{c} + 2\omega)^2 + \Gamma^2]}	\\
				&=	\frac{D^2\omega(\Gamma^2 + \omega^2 - \Omega^2)}{\Omega^2(\Gamma^2\omega^2 + (\omega^2-\Omega^2)^2)}.
 \end{split}
\end{equation}
This is identical to Eq.~(\ref{eq peaked SD im gamma}).

\end{document}